\documentclass[useAMS,usenatbib]{mn2e}
\usepackage{epsfig}

\def\msun{{\rm M_{\odot}}}

\def\me{{\dot M_{\rm Edd}}}

\def\mo{{\dot M_{\rm out}}}
\def\le{{L_{\rm Edd}}}

\title[SMBH -- Bulge Relation] {Self--Regulated Star Formation and the Black Hole -- Galaxy Bulge Relation}

\author[C. Power, K. Zubovas, S. Nayakshin, A. R. King] 
  {C. Power\thanks{E-mail: chris.power@astro.le.ac.uk}, K. Zubovas, S. Nayakshin, \& A. R. King \\ 
  Theoretical Astrophysics Group, University of Leicester, Leicester LE1 7RH, United Kingdom}

\date{\today}

\volume{000}

\pagerange{\pageref{firstpage}--\pageref{lastpage}} \pubyear{2011}

\begin{document}

\label{firstpage}

\maketitle

\begin{abstract}

We show that star formation in galaxy bulges is self--regulating through 
momentum feedback, limiting the stellar bulge mass to $M_b \propto 
\sigma^4$. Together with a black hole mass $M_{\rm BH} \propto \sigma^4$ 
set by AGN momentum feedback, this produces a linear $M_{\rm BH} - M_b$ 
relation. At low redshift this gives $M_{\rm BH}/M_b \sim 10^{-3}$, 
close to the observed ratio. We show that AGN feedback can remove any 
remaining gas from the bulge and terminate star formation once the 
central black hole reaches the $M_{\rm BH} - \sigma$ value, contrary to 
earlier claims. We find a mild upward deviation from the $\sigma^4$ law 
at higher redshift and at higher $\sigma$.

\end{abstract}

\begin{keywords}
  accretion: accretion discs -- galaxies: formation -- galaxies:
  active -- black hole physics
 
\end{keywords}

\section{Introduction}
\label{sec:introduction}

It is now widely accepted that the centre of every medium--to--large 
galaxy contains a supermassive black hole (SMBH). The mass $M$ of this 
SMBH is observed to correlate strongly with properties of the host 
galaxy in two ways. First is the correlation with the stellar velocity 
dispersion $\sigma$ of the galaxy bulge,

\begin{equation}
M_{\rm BH} \simeq 1.5\times 10^8\sigma_{200}^4\msun
\label{msigobs}
\end{equation}

\noindent \citep[cf.][]{2000ApJ...539L...9F,2000ApJ...539L..13G}, where 
$\sigma_{200} = \sigma/200~{\rm km\,s^{-1}}$. Second is the correlation 
with bulge stellar mass $M_b$

\begin{equation}
M_{\rm BH} \simeq 1.6\times 10^{-3}M_b 
\label{mm}
\end{equation}

\noindent \citep[cf. ][]{2004ApJ...604L..89H}. The first of these 
relations (\ref{msigobs}) is considerably tighter than the second 
(\ref{mm}), but both are highly significant statistically.

The existence of these relations strongly suggests a connection between 
the growth of the SMBH and of its host galaxy, despite the huge 
disparity in their masses. The underlying reason is evidently that the 
accretion energy $E_h = \epsilon_{\rm acc} Mc^2$ released in building 
the SMBH (where $\epsilon_{\rm acc} \sim 0.1$ is the energy efficiency 
of accretion) greatly exceeds the binding energy of the bulge $E_b \sim 
M_b\sigma^2$ by a factor

\begin{equation}
{E_h\over E_b} \sim {360\over \sigma_{200}^{2}}.
\label{bind}
\end{equation}

\noindent Thus absorption of even a small fraction of the accretion 
energy released by the SMBH must significantly affect the bulge 
\citep{2003ApJ...596L..27K}. This suggests that the process connecting 
SMBH growth and galaxy growth is some kind of feedback from the SMBH. 
Because galaxies are generally optically thin this feedback must be 
mechanical, and so presumably involves matter expelled during the 
accretion process. This in turn strongly suggests that feedback is a 
consequence of super--Eddington accretion. This is plausible because the 
\citet{1982MNRAS.200..115S} relation linking the growth of SMBHs to 
luminous accretion, and the low observed fraction of active galaxies 
among all galaxies \citep[e.g.][]{2004ApJ...613..109H}, together require 
that most SMBH growth occurs in such phases.

The remaining obstacle to explaining the $M_{\rm BH} - \sigma$ relation 
(\ref{msigobs}) is the very inefficient mechanical coupling of accretion 
energy implied by (\ref{bind}). This explains why the proposal by 
\citet{1998A&A...331L...1S}, that a significant fraction of a SMBH's 
Eddington energy rate

\begin{equation}
\le = {4\pi G M_{\rm BH} c\over \kappa}
\label{ledd}
\end {equation}

\noindent (where $\kappa$ is the Thomson opacity) is deposited as 
feedback into the surrounding galaxy, implies SMBH masses of order only 
$\sim 10^4 - 10^5\msun$ rather than $\sim 10^8\msun$ for a typical 
galaxy \citep[cf.][]{2010MNRAS.402.1516K}. Instead, two further features 
of the problem imply a theoretical relation very close to the observed 
one (\ref{msigobs}),

\begin{equation}
M_{\rm BH} = M_{\sigma} = {f_g\kappa\over \pi G^2}\sigma^4,
\label{msig}
\end{equation}

\noindent where $f_g \simeq 0.16$ is the baryon fraction relative to 
dark matter \citep[cf.][]{2003ApJ...596L..27K,2005ApJ...635L.121K}. 
These new features are (a) the observation that SMBH accretion is never 
highly super--Eddington, so that each emitted photon scatters on average 
only once, implying that the wind from the SMBH region carries momentum 
at the Eddington rate $\le/c$; and (b) in impacting the interstellar 
medium of the host bulge, the wind shocks and loses almost all its 
energy through Compton cooling by the AGN radiation field. As 
\citet{2010MNRAS.402.1516K} notes, feature (a) implies the we can treat 
the problem in the single--scattering limit, while feature (b) implies 
that the wind impact creates a {\it momentum--driven} outflow in the 
host ISM.\footnote{Similar arguments are
  presented in \citet{2005ApJ...618..569M}, who obtain a result 
identical to
  (\ref{msig}). However, they assume
  that it is dust rather than electron scattering that drives the 
outflow,
  which implies multiple rather than single photon scatterings. The 
resulting
  momentum flux scales as $\tau\,L/c$, where $\tau$ is the optical 
depth, and
  so a SMBH that is formally sub-Eddington in the electron scattering 
sense
  can radiate effectively at the SMBH's Eddington limit.}

It follows that, for SMBH masses $M_{\rm BH} < M_{\sigma}$, an outflow 
from the vicinity of the SMBH vicinity will remain trapped very close to 
it. Once $M_{\rm BH}$ reaches the critical value $M_{\sigma}$, the 
momentum rate $\le/c$ of the wind precisely balances the total weight 
$4f_g\sigma^4/G$ of the overlying gas (which, assuming isothermal mass 
distributions for both the dark matter and baryons, is independent of 
radial extent; cf. \citealt{2009MNRAS.398L..54N}). The equality between 
thrust and weight then gives the theoretical relation (\ref{msig}), as 
the outflow propagates through the bulge on its dynamical time and 
presumably halts further accretion on to the SMBH. Precisely how this 
last step occurs is still the most unclear part of any theory of the 
$M_{\rm BH} - \sigma$ relation \citep[see, for 
example,][]{2010MNRAS.402..789N}.\\

Self-regulated SMBH growth may explain the $M_{\rm BH} - \sigma$ 
relation, but what of the $M_{\rm BH} - M_b$ relation (\ref{mm})? As 
noted in \citet{2003ApJ...596L..27K}, Compton cooling becomes 
inefficient beyond a certain radius $R_c$ within the bulge. At a SMBH 
mass $M_{\rm BH} = M_{\sigma}$ the shock rapidly reaches this radius and 
then accelerates to a higher speed because of the thermal expansion of 
the shocked wind, which now no longer cools. Thus the outflow changes 
its character from momentum--driven to energy--driven at this point.

This high--speed outflow removes any ambient gas from the bulge, 
quenching star formation in the process, and so it is tempting to 
identify the original baryonic mass within this radius as the bulge mass 
$M_b$, as was argued by \citet{2003ApJ...596L..27K} and 
\citet{2005ApJ...635L.121K}). However, such an identification is not 
entirely satisfactory because the relation of the original gas mass to 
the expected stellar mass of the bulge in this model remains unclear -- 
it may be all or nothing. The goal of this short paper is to address 
this deficiency of the model. \\

We first recall the results of \citet{2006ApJ...650L..37M}, who pointed 
out that star formation in the nuclear star cluster (NC) of a galaxy may 
provide enough momentum thrust to establish its own $M_{\rm NC} - 
\sigma$ relation, offset from the SMBH $M_{\rm BH} - \sigma$ relation. 
They emphasised that the momentum feedback produced by young NCs is 
actually quite similar to that produced by a SMBH accreting at its 
Eddington rate but smaller by a factor of $\sim 20$ for a 
\citet{2003PASP..115..763C} IMF. This similarity arises because massive 
young stars are observed to radiate near their Eddington limit and to 
produce outflows at rates almost matching that of a black hole of the 
same mass in the model of 
\citet{2003ApJ...596L..27K,2005ApJ...635L.121K}.

This paper extends the idea of \citet{2006ApJ...650L..37M} by 
investigating momentum feedback from a galaxy's stellar bulge, rather 
than its central NC. In particular, we explore the role stellar feedback 
could play in regulating bulge mass $M_b$ and in shaping the $M_{\rm BH} 
- M_b$ relation. We note that \citet{2005ApJ...618..569M} followed a 
similar approach in their explanation of the origin of the 
\citet{1976ApJ...204..668F} relation ($L \propto \sigma^4$) for 
elliptical galaxies. There is an important distinction, however, between 
feedback from a galaxy's bulge and feedback from its NC. 
\citet{2006ApJ...650L..37M} considered {\em instantaneous} feedback from 
young massive stars in a NC while they are still on their main sequence. 
In contrast, the shortest timescale in a galaxy bulge is the dynamical 
timescale, which is much longer than the lifetimes of massive stars 
\citep[cf.][]{2009MNRAS.398L..54N}, and so we must treat feedback from 
the bulge in a time-integrated or {\em extended} form. As for NCs, the 
feedback produced in this case is proportional to the total mass of the 
bulge -- but with a smaller efficiency because the stellar population in 
the bulge is relatively mature.

We explore these ideas in more detail in the following sections. We 
begin by discussing star formation and the role it plays in regulating 
the masses of the galaxy bulges (\S~\ref{sec:star_formation}); then we 
examine how gas is swept out of the bulge by black hole feedback 
(\S~\ref{sec:swept_out}); and finally, we assess the implications of 
these ideas for the $M_{\rm BH}-M_b$ relation (\S~\ref{sec:mbhmb}).

\section{Star formation in galaxy bulges}
\label{sec:star_formation}

We begin our analysis by considering how star formation in a galaxy 
bulge impacts on the mass of virialised gas $M_{\rm g,vir}$ in a dark 
matter halo. A fraction of the virialised gas mass will be converted 
into bulge stars ($M_b$), but once the stellar mass in the bulge exceeds 
a critical value ($M_{\rm max}$), momentum-driven feedback from the 
stars (in the form of winds and supernovae) will expel this gas and help 
to regulate the bulge mass by suppressing further star formation. Our 
goal is to establish what mass of stars must form to suppress further 
star formation. As we noted in the introduction, a similar problem has 
been explored by \citet{2005ApJ...618..569M} as an explanation for the 
origin of the \citet{1976ApJ...204..668F} relation for elliptical 
galaxies, $L \propto \sigma^4$.

Massive stars dominate the mass and momentum outflow from star clusters 
through supernovae and stellar winds 
\citep[cf.][]{1992ApJ...401..596L,1999isw..book.....L}. For a 
\citet{1955ApJ...121..161S} initial mass function (IMF) and solar 
metallicity, the contributions peak at stellar masses of $12\msun$ for 
supernovae and about $50 \msun$ for momentum outflow via winds. 
Integrated over time, these two feedback mechanisms contribute roughly 
equally to the momentum outflow from massive star clusters. 
Consequently, the characteristic main sequence age for stars dominating 
momentum feedback is $t_{\rm ms} \simeq 10^7$ years. This is of order 
the dynamical time in inner parts of the dark matter halo and shorter 
than the Salpeter timescale ($t_{\rm S} \simeq 4\times 10^7$ years) on 
which the SMBH grows. This means that the rate of stellar momentum 
feedback is fixed by the rate at which new massive stars form and 
replace ones that are dying or have died \citep[see Fig. 12 in] 
[]{1992ApJ...401..596L}. In this regime, star formation injects momentum 
to the ambient medium at the global rate

\begin{equation}
\dot p_{\ast} \simeq \epsilon_{\ast}c\dot M_{\ast}
\label{pdotstar}
\end{equation} 

\noindent where $\epsilon_{\ast} \simeq 10^{-3}$ (cf. 
\citealt{1992ApJ...401..596L}; see also \S2.2 of 
\citealt{2005ApJ...618..569M}).

Hence the total momentum produced by a stellar mass $M_{\ast}$ is simply
\begin{equation}
p_{\ast} \simeq \epsilon_{\ast}cM_{\ast} .
\label{pstar}
\end{equation}

\noindent Feedback may inhibit star formation when gas that is not 
locked into stars acquires a typical velocity of order the velocity 
dispersion $\sigma$. Let $M_0$ be the initial bulge gas mass prior to 
the onset of star formation; we assume that $M_0$ is of order the virial 
value $M_{\rm g,vir}$. Feedback inhibits further star formation in the 
bulge when the total momentum injected into bulge gas reaches

\begin{equation}
  p_{\rm inh} \simeq \eta  M_0 \sigma,
\end{equation}

\noindent where $\eta \sim 1$. $\eta$ captures the uncertainty in the 
amount of momentum feedback that the system can absorb before star 
formation is suppressed. For example, if the gas needs to be completely 
expelled from the halo by star formation feedback rather than by SMBH 
feedback, then we might expect $\eta$ to be greater than unity because 
the average escape velocity from the galaxy will be typically just above 
$\sigma$. On the other hand, if SMBH expels the remaining gas then 
$\eta$ could be less than unity: star formation feedback may simply need 
to slow down collapse of gas into stars until the SMBH grows up to the 
$M_{\rm BH}-\sigma$ value 
\citep[e.g.][]{2005ApJ...618..569M,2009MNRAS.398L..54N}. Observational 
evidence can provide important guidance as to what the value of $\eta$ 
might be, as we discuss below.

The derivation of the $M_{\rm BH} - \sigma$ relation in 
\citet{2005ApJ...635L.121K} shows that mechanical feedback from 
accretion onto the SMBH is confined to a small region near it initially. 
This suggests that stellar feedback is the most plausible mechanism by 
which star formation can be inhibited up until this point \citep[see 
also \S~5 of][]{2005ApJ...618..569M}. Therefore, this means that the 
total stellar bulge mass $M_b$ cannot exceed a value $M_{\rm max}$ such 
that $p_{\ast} = p_{\rm inh}$; this means that

\begin{equation}
\epsilon_{\ast}cM_{\rm max} \simeq \eta M_0\sigma
\end{equation}
from (\ref{pstar}), which gives
\begin{equation}
  M_b \la M_{\rm max} \simeq \eta M_0 \frac{\sigma}{\epsilon_{\ast}\,c},
 \label{mass_bulge}
\end{equation}
or
\begin{equation}
  M_b \la M_{\rm max} \simeq 0.6 \eta M_0 \sigma_{200};
 \label{mass_bulge_limit}
\end{equation}
where $\sigma_{200}=\sigma/200 \rm kms^{-1}$. For the moment we assume that
$\eta=1$ but we keep in mind our discussion from above.

\section{Sweeping out the bulge}
\label{sec:swept_out}

Inspection of relation (\ref{mass_bulge_limit}) suggests that a 
significant gas mass $M_{\rm gas}$ might be present in a galaxy bulge as 
the stellar mass asymptotes towards its limiting value $M_b$; we would 
expect $M_{\rm gas}$ to be of order the baryon virial mass $M_{\rm 
g,vir}$. The observed $M_{\rm BH} - M_b$ relation then implies that 
feedback from the central SMBH must be able to sweep away this gas as 
$M_{\rm BH}$ reaches the value $M_{\sigma}$ given by the $M_{\rm BH} - 
\sigma$ relation (\ref{msig}). As the escape speed from an isothermal 
bulge is $\simeq 2\sigma$, the SMBH must supply an energy

\begin{equation}
E_{\rm esc} = {1\over 2}M_{\rm gas}(2\sigma)^2
\label{esweep}
\end{equation}
to this gas.

To check this, we assume that the wind driven by super--Eddington 
accretion on to the SMBH provides thrust given by the single--scattering 
limit, i.e.

\begin{equation}
\mo v \simeq {\le\over c},
\label{ss}
\end{equation}

where $\mo$ is the wind mass outflow rate and $v$ is its speed. This 
limit implies that the wind momentum is similar to the original photon 
momentum, i.e. the scattering optical depth is $\sim 1$ and each photon 
scatters on average about once before escaping 
\citep[cf.][]{1999isw..book.....L}. This is likely to hold for SMBH 
accretion because even the dynamical accretion rate is not much larger 
than the Eddington rate $\me$ for the hole near its $M - \sigma$ mass. 
That this is so is shown explicitly in \citet{2010MNRAS.402.1516K}, 
which also shows that \begin{equation} v \simeq \epsilon_{\rm acc} c 
\end{equation} in these circumstances, where $\epsilon_{\rm acc} \simeq 
0.1$ is the accretion efficiency. Accordingly the total energy of the 
outflow is 

\begin{equation} 
E_{\rm out} = {1\over 2}\Delta M_{\rm BH}v^2 = {1\over 2}\Delta M_{\rm
  BH}(\epsilon_{\rm acc} c)^2,
\end{equation}

where $\Delta M_{\rm BH}$ is the increase in SMBH mass during the 
accretion episode considered. The fraction of $E_{\rm out}$ supplied to 
the bulge gas now depends on the nature of the shock as the wind impacts 
on it. If the shocked gas cools rapidly (faster than the flow time), the 
outflow is {\it momentum--driven} \citep[e.g.][]{1997pism.book.....D} 
and only a small fraction of $E_{\rm out}$ is used to drive the outflow. 
However, this turns out to be slightly too small to expel it (as we show 
in the appendix). In contrast, if the shocked gas does not readily cool 
(an {\it energy--driven} outflow), the thermal expansion of the shocked 
gas means that almost all of $E_{\rm out}$ is transferred to the gas to 
expel it.

\citet{2003ApJ...596L..27K,2005ApJ...635L.121K} has shown that the 
outflow is momentum--driven when very close to the SMBH, but 
energy--driven outside a radius $\sim\!1$~kpc for a typical galaxy 
bulge\footnote{The precise value of this radius scales as $(v/c)^2$. 
This can be seen by comparing the flow time of the outflow to the 
Compton cooling time in the shock preceeding the outflow, given by 
equations (9) and (8) respectively of \citet{2003ApJ...596L..27K}, with 
equation (14) for the flow velocity corrected by a factor of 2 to be 
consistent with the value of $M_{\rm BH}$ derived in 
\citet{2005ApJ...635L.121K}. In our analysis we assume a wind velocity 
of $\sim\!0.1\,c$, which is consistent with observational data
  \citep[cf. Figure 8 of][]{2010A&A...521A..57T}.}. This is because the 
efficiency with which the quasar radiation field Compton cools the shock 
preceeding the outflow declines with increasing radius 
\citep[cf.][]{1997ApJ...487L.105C}. If we require that the thrust of the 
momentum-driven
outflow must exceed the weight of the overlying gas, as must be the case if the outflow is to escape
from the immediate vicinity of the black hole, then it is straightforward to show that a 
$M_{\rm BH} - \sigma$ of the form given by equation (\ref{msig}) must follow. 

However the total gas mass that the outflow can ultimately remove is 
given by considering the total energy $E_{\rm out}$, as the outflow 
rapidly becomes energy--driven. By equating $E_{\rm out}$ and $E_{\rm 
esc}$ we find that

\begin{equation}
{\Delta M_{\rm BH}\over M_{\rm gas}} = \left({2\sigma\over \epsilon_{\rm acc} c}\right)^2 
= 1.8\times 10^{-4}\sigma_{200}^2
\label{deltam}
\end{equation}

i.e. only a small increase in SMBH mass is needed to sweep the bulge 
clear of any remaining gas. We quantify this further below.

\section{The Black Hole -- Bulge Mass Relation}
\label{sec:mbhmb}

We have argued that momentum--driven outflows from star formation tend 
to produce a bulge stellar mass

\begin{equation}
{M_b \la 0.6\eta\,M_{\rm g,vir} \sigma_{200}.}
\label{mbulge}
\end{equation}

Here we assume that $M_0$ in equation (\ref{mass_bulge_limit}) is of 
order the virialised gas mass within the dark matter halo, i.e. $M_{\rm 
g,vir} = f_g\,M_v$, where $M_v$ is the total virial mass, and $\eta$ 
captures the uncertainty in the amount of momentum feedback the system 
can absorb before star formation is suppressed. For the sake of 
simplicity, we also assume that both the gas and dark matter follow 
isothermal mass distributions, and that the velocity dispersion $\sigma$ 
of the bulge and its underlying dark matter halo are the same. Neither 
of these assumptions are likely hold in detail, but their effect is 
quantitative rather than qualitative. Assuming isothermality, how does 
the virial gas mass $M_{\rm g,vir}$ vary with velocity dispersion 
$\sigma$?

We make the standard assumption that matter is virialised within a 
radius such that the mean density is 200 times the critical value; for 
the particular case of an isothermal sphere, this gives

\begin{equation}
  R_v = {\sigma\over 5 \sqrt{2} H}
\end{equation} 

where $H(z)=100\,h(z)\rm kms^{-1} Mpc^{-1}$ is the Hubble parameter at 
the given redshift $z$, with $h(z)$ the dimensionless Hubble parameter. 
Because $M_v = 2\sigma^2R_v/G$ we find that

\begin{equation}
  M_b \la 5.1 \times 10^{11}\,\eta\,{f_g\over 0.16}{\sigma_{200}^4\over 
    h(z)}\msun;
\end{equation}

when combined with equation (\ref{msig}), we find that

\begin{equation}
  {M_{\sigma}\over M_b} \ga {7.3\times 10^{-4}\,h(z) \over \eta}.
  \label{mm2}
\end{equation}

in reasonable agreement with, for example, \citet{2004ApJ...604L..89H}.

We expect deviations from this relation and the simple $M_{\rm BH} \sim 
\sigma^4$ law, as $M_{\rm BH}$ may grow above $M_{\sigma}$ by an amount 
$\la \Delta M_{\rm BH}$ in expelling any remaining gas. Combining 
(\ref{mm2}) with (\ref{deltam}) -- where we assume that $M_{\rm gas}$ is 
equal to $M_{\rm g,vir}$ -- and (\ref{mbulge}) gives the predicted 
maximum deviation from the theoretical relation

\begin{equation}
  {\Delta M_{\rm BH}\over M_{\sigma}} \la 0.41\,{\sigma_{200}\over h(z)}.
\end{equation}

This implies that $M_{\rm BH}$ should tend to increase above 
$M_{\sigma}$ with redshift and with $\sigma$. With $M_{\rm BH} = 
M_{\sigma} + \Delta M_{\rm BH}$ the black hole -- bulge mass relation 
finally becomes

\begin{equation}
  {M_{\rm BH}\over M_b} \sim {7.3\times 10^{-4}\,h(z)\over \eta}\left[1 +
    {0.41\sigma_{200}\over h(z)}\right] 
  \label{mm3}
\end{equation}

Assuming $h=0.7$ at $z=0$, this implies that
\begin{equation}
  {M_{\rm BH}\over M_b} \sim 1.15\times 10^{-3} \eta^{-1}
\end{equation}
for a galaxy with $\sigma_{200} = 1$, close to the observed relation (\ref{mm}).
This relation should curve slightly upwards for larger $\sigma$. 

\section{Discussion}
\label{sec:discussion}

We have assumed that star formation in galaxy bulges is self--regulating 
through momentum feedback acting on gas in the galaxy's dark matter 
halo. This puts a limit on the total mass of the stars that can form 
within the halo (equation \ref{mbulge}). Crucially, this maximum bulge 
mass scales as $\sigma^4$ with the galaxy velocity dispersion. Given 
that the SMBH mass also scales as $\sigma^4$ 
\citep[cf.][]{2003ApJ...596L..27K,2005ApJ...635L.121K,2005ApJ...618..569M}, 
this results in a linear $M_{\rm BH} - M_b$ relation, close to the 
observed one \citep{2004ApJ...604L..89H}. This relation arises because 
both the SMBH and the bulge are limited by essentially the same quantity 
-- the maximum momentum thrust that the system can take before the gas 
is blown away.

The best match to the observed relation \citep{2004ApJ...604L..89H} is 
obtained for the dimensionless parameter $\eta \sim 0.7$. In the context 
of our order of magnitude derivation in \S \ref{sec:star_formation}, a 
value of $\eta$ less than one implies that the bulge is not entirely 
self-sufficient in limiting its own growth, and that a SMBH or a NC are 
required to terminate bulge growth completely.

Finally, we argued in \S \ref{sec:mbhmb} that if the SMBH growth 
timescale (few $\times$ Salpeter) is shorter than the star formation 
(dynamical) timescale, the growth of the central black hole is able to 
shut off star formation in the bulge early. In this case the extra SMBH 
growth introduces a mild upward trend in the $M_{\rm BH} \propto 
\sigma^4$ relation.

\section{Acknowledgments}

The authors thank the anonymous referee for their comments that helped 
to improve the clarity of the manuscript. CP and KZ acknowledge support 
via an STFC theoretical astrophysics rolling grant and an STFC 
studentship respectively.

\section*{Appendix}

If one assumes that the outflow is momentum--driven at all radii (i.e. 
efficient shock cooling everywhere), the outflow would exert a thrust given
precisely by the weight $4f_g\sigma^4/G$ of the overlying gas of mass $M_{\rm gas}$. 
Then the total work done by this thrust on the gas, assumed to extend to
some radius $R$, is 
\begin{equation}
E'_{out} = {4f_g\sigma^4R\over G} = 2\sigma^2M_{\rm gas} = {1\over 2}E_{\rm esc}
\end{equation}
\noindent because the gas is in equilibrium with the isothermal dark matter
halo. Therefore a purely momentum--driven outflow would fail by a factor 2
to remove the overlying gas, which is essentially the result claimed by
\citet{2010ApJ...725..556S}. 

However, cooling is ineffective once the shock radius is $\ga 1$~kpc for a 
typical galaxy and so the outflow becomes energy--driven. This is because the
shock is Compton cooled while it is close to the SMBH and the radius at which
the Compton cooling time begins to exceed the shock flow time varies as 
$0.75 \sigma_{200}^3$~kpc, where where we've assumed a wind velocity of
$0.1c$. If the shock is to cool radiatively, it must do so by means of thermal 
bremsstrahlung because of the magnitude of the shock temperature 
($T \simeq 10^{12}(v/c)^2 K \sim 10^{10}\rm K$ for $v=0.1 c$). However, the 
free-free cooling time is long in this regime -- $\sim 5 \times 10^8 n^{-1}$ 
years where $n$ is the electron number density -- and it exceeds the shock 
flow time at all interesting radii. For this reason, once the shell becomes 
energy-driven, it remains energy driven.

\label{lastpage}

\end{document}